\documentclass[showpacs,preprint]{revtex4}
\usepackage{amsmath,amssymb,mathrsfs}
\usepackage{latexsym}
\usepackage{graphicx,psfrag} 

\newcommand{\half}{$\frac{1}{2}$ }

\newcommand{\ket}[1]{\left|#1\right\rangle}

\newcommand{\up}{\uparrow}
\newcommand{\dw}{\downarrow}
\def\bd{\begin{displaymath}}
\def\ed{\end{displaymath}}
\def\be{\begin{equation}}
\def\ee{\end{equation}}
\def\bea{\begin{eqnarray}}
\def\eea{\end{eqnarray}}
\def\bi{\begin{itemize}}
\def\ei{\end{itemize}}
\def\bn{\begin{enumerate}}
\def\en{\end{enumerate}}

\def\ie{{\it i.e.},\ }

\allowdisplaybreaks[1]
\begin{document}
\title{Exact two-holon wave functions in the 
Kuramoto--Yokoyama model}
\author{Ronny Thomale, Dirk Schuricht, and Martin Greiter}
\affiliation{Institut f\"ur Theorie der Kondensierten Materie,\\
  Universit\"at Karlsruhe, Postfach 6980, 76128 Karlsruhe}
\pagestyle{plain}
\begin{abstract}
  We construct the explicit two-holon eigenstates of the SU(2)
  Kuramoto--Yokoyama model at the level of explicit wave functions. We derive
  the exact energies and obtain the individual holon momenta, which are
  quantized according to the half-Fermi statistics of the holons.
\end{abstract}
\pacs{75.10.Pq, 02.30.Ik, 05.30.Pr, 75.10.Jm}

\maketitle

\section{Introduction}

One milestone towards the understanding of fractional quantization in
one dimension is the $1/r^2$ model independently introduced by
Haldane~\cite{Haldane88} and Shastry~\cite{Shastry88} in 1988. The
model describes a spin 1/2 chain with a Heisenberg interaction which
falls off as one over the square of the distance between the sites.
The exact ground state is provided by a trial wave function proposed
by Gutzwiller~\cite{Gutzwiller63} as early as in 1963.  The
Haldane--Shastry Model (HSM) offers the opportunity of studying
spinons, \ie the elementary excitations of one-dimensional spin
chains, on the level of explicit and analytical expressions for one and
two-spinon wave functions~\cite{BGLtwospinon}, which are at least at
present not available for any other model.  Kuramoto and
Yokoyama~\cite{KuramotoYokoyama} generalized the model to allow for
mobile holes (\ie empty lattice sites) with a hopping parameter that
also falls off with $1/r^2$ as a function of the distance.  The
Kuramoto--Yokoyama Model (KYM) hence contains spin and charge degrees
of freedom, and accordingly supports spinon and holon excitations,
which carry spin \half but no charge and charge $+1$ but no spin,
respectively.  In principle, the KYM allows for a similarly
explicit construction of holon wave functions, which so far have only
been obtained for states involving a single holon.  The reason for
this deficit has been of technical nature, related to the commutation
relations of the operators used to build the Hilbert space of these
fractionally quantized excitations.  Whereas for the spinons one can
use bosonic spin-flip operators, one needs fermionic creation and
annihilation operators for the holons.

In this article we address and overcome this technical problem as we
construct the explicit wave functions for two-holon excitations of the
KYM.  The article is organized as follows:  In Section~\ref{seckyh} we
review the KYM and its properties. In Section~\ref{secvcs}
and~\ref{secspe}, we briefly discuss the ground state at half filling
and the spinon excitations.  We further review the analytic results so
far known for the one-holon excitations in Section~\ref{sec1he} as a
preliminary for the construction of the explicit two-holon wave
functions to be done in Section~\ref{sec2he}.  Therein we derive the
exact energies and individual holon momenta, which turn out to be
quantized according to half-Fermi statistics of the holons.

\section{Kuramoto--Yokoyama model}\label{seckyh}

The Kuramoto--Yokoyama model~\cite{KuramotoYokoyama} is most
conveniently formulated by embedding the one-dimensional chain with periodic
boundary conditions into the complex plane by mapping it onto the unit circle
with the sites located at complex positions
$\eta_\alpha=\exp\!\left(i\frac{2\pi}{N}\alpha\right)$, where $N$ denotes the
number of sites and $\alpha=1,\ldots,N$.  The sites can be either singly occupied by an up or down-spin electron or empty.  The Hamiltonian is given
by
\begin{equation}
H_{\text{KY}}=-\frac{\pi^2}{N^2}
\sum_{\substack{\alpha,\beta=1 \\\alpha \neq \beta}}^{N}
\frac{P_{\alpha \beta}}{\vert \eta_{\alpha} - \eta_{\beta} \vert^2},
\label{ky1}
\end{equation}
where $P_{\alpha \beta}$ exchanges the configurations on the sites
$\eta_{\alpha}$ and $\eta_{\beta}$ including a minus sign if both are
fermionic. Rewriting \eqref{ky1} in terms of spin and electron creation and
annihilation operators yields
\begin{equation}
H_{\text{KY}} =  \frac{2\pi^2}{N^2} 
\sum_{\alpha \neq \beta}^N \frac{1}{ | \eta_\alpha - \eta_\beta |^2 } 
P_{\text{G}} \Biggl[-\frac{1}{2} 
\sum_{\sigma=\up\dw}\Bigl(c_{\alpha \sigma}^\dagger 
c_{\beta \sigma}^{\phantom{\dagger}} 
+ c_{\beta \sigma}^\dagger c_{\alpha \sigma}^{\phantom{\dagger}} \Bigr)
+\vec{S}_\alpha \cdot \vec{S}_\beta
-\frac{n_{\alpha}n_{\beta}}{4}+n_{\alpha}-\frac{1}{2}\Biggr] P_{\text{G}},
\label{ky2}
\end{equation}
\noindent
where the Gutzwiller projector $P_{\text{G}} = \prod_\alpha ( 1 - c_{\alpha
  \uparrow}^\dagger c_{\alpha \downarrow}^\dagger c_{\alpha
  \downarrow}^{\phantom{\dagger}} c_{\alpha \uparrow}^{\phantom{\dagger}} )$
enforces at most single occupancy on all sites. The charge occupation and spin
operators are given by $n_\alpha = c_{\alpha \uparrow}^\dagger c_{\alpha
  \uparrow}^{\phantom{\dagger}} + c_{\alpha \downarrow}^\dagger c_{\alpha
  \downarrow}^{\phantom{\dagger}}$ and $S_\alpha^a = \frac{1}{2} \sum_{\sigma
  , \sigma^{'}} c_{\alpha \sigma}^\dagger \tau^a_{\sigma \sigma^{'}} c_{\alpha
  \sigma^{'}}$, where $\tau^a$, $a=x,y,z$, denote the Pauli matrices.

The interaction strength in \eqref{ky1} is an analytic function of the lattice
sites by use of
\begin{equation}
\frac{1}{\vert \eta_{\alpha}-\eta_{\beta}\vert^2}=
-\frac{\eta_{\alpha}\eta_{\beta}}{(\eta_{\alpha}-\eta_{\beta})^2}.
\label{eq:vertetavert}
\end{equation}

The KYM is supersymmetric, \ie the Hamiltonian \eqref{ky1} commutes with the
operators $J^{ab}=\sum_\alpha a_{\alpha a}^{\dagger}a_{\alpha
  b}^{\phantom{\dagger}}$, where $a_{\alpha a}$ denotes the annihilation
operator of a particle of species $a$ ($a$ runs over up- and down-spin as well
as empty site) at site $\eta_\alpha$. The traceless parts of the operators
$J^{ab}$ generate the Lie superalgebra su(1$|$2), which includes in particular
the total spin $\vec{S}=\sum_{\alpha=1}^N\vec{S}_\alpha$.  In
addition, the KYM possesses a super-Yangian symmetry~\cite{Haldane94}, which
causes its amenability to rather explicit solution.

\section{Vacuum State}\label{secvcs}

We first review the ground state at half filling, which is the state
containing no excitations (neither spinons nor holons). For $N$ even,
this vacuum state is constructed by the Gutzwiller projection of a filled
band (or Slater determinant (SD) state) containing a total of $N$
electrons:
\begin{equation}
\ket{\Psi_0}=P_{\text{G}}\prod_{\vert q \vert < q_{\text{F}}}
c_{q \uparrow}^{\dagger}c_{q \downarrow}^{\dagger}\ket{0}\equiv 
P_{\text{G}}\ket{\Psi_{\text{SD}}^{N}}. \label{su2vac}
\end{equation}
Taking the fully polarized state $\ket{0_{\downarrow}}=\prod_\alpha c_{\alpha
  \downarrow}^{\dagger}\ket{0}$ as reference state, we can rewrite the vacuum
state as
\begin{equation}
\ket{\Psi_0}=\sum_{\{z_i\}}\Psi_0(z_1,\ldots,z_M)\, 
S^{+}_{z_1}\dots S^{+}_{z_M}\ket{0_{\downarrow}}, 
\label{groundstate}
\end{equation}
where $M=N/2$ and the $z_i$'s denote the up-spin coordinates. The sum in
(\ref{groundstate}) extends over all possible ways to distribute the
coordinates $z_i$'s over the lattice sites $\eta_\alpha$.  The wave function
is given by~\cite{Haldane88,Shastry88}
\begin{equation}
\Psi_0(z_1,\ldots,z_M)=\prod_{i < j}^M (z_i-z_j)^2 \prod_{i=1}^M z_i,
\label{gswave}
\end{equation}
its energy is
\begin{equation}
E_0=-\frac{\pi^2}{4N}. \label{vacen}
\end{equation}
The total momentum of a state is evaluated by considering the operator
$\mathbf{T}$, which translates all coordinates counterclockwise by one site.
$\mathbf{T}$ is related to the momentum operator $\mathbf{P}$ via
\begin{equation}
\mathbf{T}=\exp (-i\mathbf{P}).
\label{eq:su2-momentumdefinition}
\end{equation}

This yields the momentum of $\ket{\Psi_0}$ to equal zero if $N$ is divisible
by four and $\pi$ otherwise.

Note that \eqref{gswave} represents the ground state of \eqref{ky1}
only at half filling, \ie when all sites are occupied. As was shown by
Kuramoto and Yokoyama~\cite{KuramotoYokoyama}, the ground state away
from half-filling can be constructed by Gutzwiller projection similar
to~\eqref{su2vac}.

\section{Spinon excitations}\label{secspe}

Let $N$ be odd and $M=(N-1)/2$. A localized spinon at site "$\eta_\gamma$" is
constructed by the Gutzwiller projection of an electron inserted in a Slater
determinant state of $N+1$ electrons:
\begin{equation}
\ket{\Psi_{\gamma}^{\text{sp}}}=
P_{\text{G}}\,c_{\gamma\downarrow}\ket{\Psi_{\text{SD}}^{N+1}}.
\label{eq:locspinon}
\end{equation}
The annihilation of the electron causes an inhomogeneity in the spin
and charge degree of freedom. After the projection, however, only the
inhomogeneity in the spin survives.  The spinon hence possesses spin
one-half but no charge. The wave function of a localized spinon is
given by~\cite{Haldane91}
\begin{equation}
\Psi_{\gamma}^{\text{sp}}(z_1,\dots,z_M)=
\prod_{i=1}^{M}(\eta_\gamma-z_i)\,\Psi_0(z_1,\ldots,z_M),\label{spinonwf}
\end{equation} 
where $\Psi_0$ is defined in~\eqref{groundstate}. Fourier transformation
yields the momentum eigenstates
\begin{equation}
\ket{\Psi_m^{\text{sp}}}=\sum_{\alpha=1}^N
(\bar{\eta}_\gamma)^{m}\ket{\Psi_\gamma^{\text{sp}}},
\label{eq:momspinon}
\end{equation}
which vanish identically unless $0\leq m \leq M$. In particular, this
implies that the localized one-spinon states \eqref{eq:locspinon} form
an overcomplete set. It is hence not possible to interpret the
``coordinate'' $\eta_\gamma$ literally as the position of the spinon.
The momentum eigenstates \eqref{eq:momspinon} are found to be exact
energy eigenstates of the KYM, with its energies given
by~\cite{Haldane91}
\begin{equation}
E_m^{\text{sp}}=\frac{2\pi^2}{N^2}\left(\frac{N-1}{2}-m\right)m.
\end{equation}
The spinons obey half-Fermi statistics, which was first found by the
investigation of their state counting rules~\cite{Haldane91prl2}.
Later it became apparent that the fractional statistics of the spinons
manifests itself in the quantization rules for the individual spinon
momenta as well~\cite{GreiterS05,Greiter}.

\section{One-Holon excitations}\label{sec1he}

The charged elementary excitations of the model are holons, the 
concept of which must be invoked whenever holes and thereby charge 
carries are doped into the chain.  A localized
holon at lattice site $\eta_\xi$ is constructed as
\begin{equation}
\ket{\Psi_\xi^{\text{ho}}}=c_{\xi\downarrow}^{\phantom{\dagger}}
P_{\text{G}}\,c_{\xi\downarrow}^{\dagger}\ket{\Psi_{\text{SD}}^{N-1}}.
\label{1hstate}
\end{equation} 
 
(Alternatively we could use the operators
$c_{\xi\up}^{\phantom{\dagger}}$ and $c_{\xi\up}^{\dagger}$.)
Compared to the spinon we eliminate the inhomogeneity in spin while
creating an inhomogeneity in the charge distribution after
Gutzwiller projection.  Thus the holon has no spin but charge $e>0$ (as
the electron charge at site $\eta_\xi$ is removed).  Note that
the holon is strictly localized at the holon coordinate $\xi$, as
holon states on neighboring coordinates are orthogonal.  In total,
there are $N$ independent one-holon states \eqref{1hstate}.

Momentum eigenstates are constructed from \eqref{1hstate} by Fourier
transformation. It turns out that only $(N+3)/2$ of them are energy
eigenstates~\cite{BGLonespinononeholon}.  We will restrict ourselves to this
subset in the following.  These states are readily described in terms of
their wave functions.  We take $\ket{0_{\downarrow}}$ as reference
state, and write the one-holon energy eigenstates
as~\cite{BGLonespinononeholon}
\begin{equation}
\ket{\Psi_m^{\text{ho}}}=\sum_{\{z_i;h\}}
\Psi_{m}^{\text{ho}} (z_1,\ldots,z_M;h)\,c_{h\downarrow}^{\phantom{\dagger}}\,
S^+_{z_1}\dots S^+_{z_M}
\ket{0_{\downarrow}}, 
\label{1hrefstate}
\end{equation} 
where the sum extends over all possible ways to distribute the up-spin
coordinates $z_i$ and the holon coordinate $h$ over the lattice sites
$\eta_\alpha$ subject to the restriction $z_i\neq h$.  The one-holon
wave function is given by
\begin{equation}
\Psi_{m}^{\text{ho}} (z_1,\ldots,z_M;h)=h^{m}\prod_{i=1}^{M}(h-z_i)\,
\Psi_0(z_1,\ldots,z_M),
\label{1hwave}
\end{equation}
where $\Psi_0$ is given by \eqref{gswave}.  Note that as a sum over
the coordinates $h$ is included in \eqref{1hrefstate}, no such sum is
required in \eqref{1hwave}.  It can be shown that $0\leq m \leq M+1$,
where $M=(N-1)/2$ is the number of up-spin coordinates, the wave
function \eqref{1hwave} represents an exact energy eigenstate with
energy~\cite{BGLonespinononeholon}
\begin{equation}
E_{m}=\frac{2\pi^2}{N^2}\,\left(m-\frac{N+1}{2}\right)\,m.
\label{1hen}
\end{equation}
For other values of $m$, the states $\ket{\Psi_m^{\text{ho}}}$ do not
vanish identically (as $\ket{\Psi_m^{\text{sp}}}$ for spinons do), but
are not eigenstates of the Kuramoto--Yokoyama Hamiltonian \eqref{ky1}
either.  Consequently, we are allowed to refer to the states
\eqref{1hrefstate} with \eqref{1hwave} as ``holons'' only if $0\leq m \leq
M+1$.

Note that this implies that the states \eqref{1hstate} do not really
constitute ``holons'' localized in position space, but only basis
states which can be used to construct holons if the momentum is chosen
adequately.  The total number of single-holon states is given by
$M+2$, according to the number of distinct values $m$ is allowed to
assume.  Since the states \eqref{1hstate} are orthogonal for different
lattice positions $\xi$, there are $N=2M+1$ orthogonal position basis
states $\ket{\Psi_\xi^{\text{ho}}}$. Hence the states
$\ket{\Psi_\xi^{\text{ho}}}$ cannot strictly be holons, but rather
constitute incoherent superpositions of holons and other states.  It
is clear from these considerations that it is not possible to localize
a holon onto a single lattice site.  The best we can do is to take a
Fourier transform of the exact eigenstates $\ket{\Psi_m^{\text{ho}}}$
for $0\leq m \leq M+1$ back into position space.  The resulting
``localized'' holon states will be true holons but will not be
localized strictly onto lattice sites.  Such a true holon state
``localized'' at a given lattice site will not be orthogonal to such a
state ``localized'' at the neighboring lattice site, as there are only
$M+2$ holon states while there are $N$ lattice sites.  The situation
is hence very similar to the case of the spinons, which form an
overcomplete set and are well known to be non-orthogonal if
``localized'' on neighboring lattice sites.

The one-holon momenta of the states \eqref{1hrefstate} with
\eqref{1hwave} are derived in analogy to the vacuum state to be
\begin{equation}
p_m^{\text{ho}}=\frac{\pi}{2}N+\frac{2\pi}{N}\left(m-\frac{1}{4}\right) 
\ \ \text{mod}\ 2\pi. 
\label{1hmom}
\end{equation} 
If we introduce the one-holon dispersion
\begin{equation}
\epsilon^{\text{ho}}(p)=
-\frac{1}{2}\left(\frac{\pi^2}{4}-p^2\right)-\frac{\pi^2}{8N^2},\quad
-\frac{\pi}{2}\le p\le\frac{\pi}{2},
\label{1hdisp}
\end{equation}
we can rewrite \eqref{1hen} with the vacuum energy \eqref{vacen} as
\begin{equation}
E_m=E_0+\epsilon^{\text{ho}}(p_m^{\text{ho}}).
\end{equation}

\section{Two-Holon excitations}\label{sec2he}

\subsection{Momentum eigenstates}

Let $N$ be even and $M=(N-2)/2$. The two-holon state with holons localized at
$\eta_{\xi_1}$ and $\eta_{\xi_2}$ is constructed in analogy to
\eqref{1hstate} as
\begin{equation}
\ket{\Psi_{\xi_1\xi_2}^{\text{ho}}}=
c_{\xi_1\downarrow}^{\phantom{\dagger}}
c_{\xi_2 \downarrow}^{\phantom{\dagger}}P_\text{G}\,
c_{\xi_1 \downarrow}^{\dagger}c_{\xi_2 \downarrow}^{\dagger}
\ket{\Psi_{\text{SD}}^{N-2}}. 
\label{2holonconstruction}
\end{equation}

In analogy to \eqref{1hwave}, a momentum basis for the two-holon 
eigenstates is provided by the wave functions
\begin{equation}
\Psi_{mn}^{\text{ho}}(z_1,\dots,z_M;h_1,h_2)=
(h_1-h_2)(h_1^m h_2^n+h_1^n h_2^m)\prod_{i=1}^M
(h_1-z_i)(h_2-z_i)\Psi_0(z_1,\dots,z_M), 
\label{twoholewave}
\end{equation}
where $\Psi_0$ is again given by \eqref{gswave}, $h_{1,2}$ denote the
holon coordinates, and the integers $m$ and $n$ satisfy
\begin{equation}
0 \leq n \leq m \leq M+1. \label{restriction}
\end{equation}
The corresponding state is then given by
\begin{equation}
\ket{\Psi_{mn}^{\text{ho}}}=\sum_{\{z_i;h_1,h_2\}}
\Psi_{mn}^{\text{ho}} (z_1,\ldots,z_M;h_1,h_2)\,
c_{h_1\downarrow}^{\phantom{\dagger}}\,c_{h_2\downarrow}^{\phantom{\dagger}}\,
S^+_{z_1}\dots S^+_{z_M}
\ket{0_{\downarrow}},
\label{2hrefstate}
\end{equation} 
where the sum extends over all possible ways to distribute the up-spin
coordinates $z_i$ and the holon coordinates $h_{1,2}$ over the lattice sites
$\eta_\alpha$ subject to the restriction $z_i\neq h_{1}\neq h_{2}$.  The momentum of the
states \eqref{2hrefstate} is easily found to be
\begin{equation}
p_{mn}^{\text{ho}}=\frac{\pi}{2}N+\frac{2\pi}{N} \bigl( m+n \bigr) \ \ 
\text{mod} \ 2\pi.
\label{su2twomomentum}
\end{equation}
It can further be shown that the states \eqref{2hrefstate} are spin singlets,
\ie they are annihilated by $S^\pm$ as well as $S^z$.

\subsection{Action of $H_{\text{KY}}$ on the momentum eigenstates}

In the following we will construct the two-holon energy eigenstates starting
from \eqref{twoholewave}.  First, we introduce the auxiliary wave
functions 
\begin{equation}
\varphi_{mn}(z_{1},\ldots,z_{M};h_1,h_2)=
h_1^{m}h_2^{n}\prod_{i=1}^M (h_1-z_i)(h_2-z_i)
\,\Psi_0(z_1,\ldots,z_M).
\end{equation}
The action of the Hamiltonian on the states~\eqref{twoholewave}
will be obtained later via
\begin{equation}
\Psi_{mn}^{\text{ho}}=\varphi_{m+1,n}+\varphi_{n+1,m}-
\varphi_{m,n+1}-\varphi_{n,m+1}.
\label{eq:auxzerlegung}
\end{equation}
Second, we rewrite the Hamiltonian~\eqref{ky2} in analogy to~\cite{BGLonespinononeholon}
as
\begin{equation}
H_{\text{KY}}=\frac{2\pi^2}{N^2}\left(H_{\text{S}}^{\text{ex}}+
H_{\text{S}}^{\text{Is}}+H_{\text{V}}+H_{\text{C}}^{\uparrow}+H_{\text{C}}^{\downarrow}\right),
\end{equation}
where we separate the spin-exchange, spin-Ising, potential, $\uparrow$-charge
kinetic term, and $\downarrow$-charge kinetic terms. In the
following we treat each term separately.

For the spin-exchange term we begin by observing that $[S_\alpha^+ S_\beta^-
\varphi_{nm}] (z_1 ,\dots , z_{M};h_1,h_2)$ is identically zero unless one of the
arguments $z_1 , ... , z_{M}$ equals $\eta_\alpha$. We have
\begin{eqnarray}
& &\hspace{-10mm}\Big[H_{\text{S}}^{\text{ex}}\varphi_{mn}\Big]
(z_{1},\ldots,z_{M};h_1,h_2)\equiv
\left[ \sum_{\alpha\neq\beta}^N
\frac{P_{\text{G}}S_{\alpha}^{+}S_{\beta}^{-}P_{\text{G}}}
{\vert \eta_\alpha - \eta_\beta \vert^2}
\varphi_{mn}\right]\!(z_{1},\ldots,z_{M};h_1,h_2) \nonumber \\
&=&\sum_{j=1}^M\sum_{\beta \neq j}^N
\frac{\eta_\beta}{\vert z_j-\eta_\beta\vert^2}
\frac{\varphi_{nm} (z_{1}, \ldots , z_{j-1}, \eta_\beta ,z_{j+1}, \ldots , z_{M};h_1, h_2)}
{\eta_\beta} \nonumber\\
&=&\sum_{j=1}^M \sum_{l = 0}^{N-1}\sum_{\beta \neq j}^N
\frac{\eta_\beta(\eta_\beta - z_j)^l}
{l!\vert z_j - \eta_\beta\vert^2}
\frac{\partial^l}{\partial z_j^l}
\left( \frac{ \varphi_{nm}(z_{1},\ldots,z_{M};h_1,h_2)}{z_j} \right) \nonumber \\
&=&\sum_{j=1}^{M}\sum_{l=0}^{N-1}A_l \frac{z_j^{l+1}}{l!}\frac{\partial^l}{\partial z_j^l}\frac{\varphi_{mn}}{z_j}\nonumber \\ 
&=&\sum_{j=1}^{M} \left( \frac{(N-1)(N-5)}{12} z_{j}
- \frac{N-3}{2} z_{j}^2 \frac{\partial}{\partial z_{j}}+ \frac{1}{2} z_{j}^3
\frac{\partial^2} {\partial z_{j}^2} \right)
 \frac{ \varphi_{mn}}{ z_{j}}\nonumber\\
&=&\left\{ \frac{M}{12}(5-2N)h_1^mh_2^n-h_1^mh_2^n\sum_{i\neq j}^{M}
\frac{1}{|z_{i}-z_{j}|^{2}}-h_1^m h_2^{n+2}
\frac{\partial^2}{\partial h_2^2}-h_1^{m+2}h_2^n
\frac{\partial^2}{\partial h_1^2} \right. \nonumber \\*
& &\phantom{,,}+\frac{N-3}{2} \left(h_1^m h_2^{n+1}
\frac{\partial}{\partial h_2}+h_1^{m+1}h_2^n
\frac{\partial}{\partial h_1}\right)+\frac{h_1^mh_2^{n+2}}{h_1-h_2}
\frac{\partial}{\partial h_2}-\frac{h_1^{m+2}h_2^n}{h_1-h_2}
\frac{\partial}{\partial h_1}\Bigg \} \frac{\varphi_{mn}}{h_1^mh_2^n},
\label{su2twoex}
\end{eqnarray}
where we have introduced the coefficients $A_l=-\sum_{\alpha=1}^{N-1}
\eta_\alpha^2 (\eta_\alpha -1)^{l-2}$.  Evaluation of the latter yields
$A_0=(N-1)(N-5)/12$, $A_1=-(N-3)/2$, $A_2=1$, and $A_l=0$ for $2<l\le
N-1$~\cite{BGLtwospinon}.

For the spin-Ising term we obtain
\begin{eqnarray}
& &\hspace{-10mm}\Big[H_{\text{S}}^{\text{Is}}\varphi_{mn}\Big]
(z_{1},\ldots,z_{M};h_1,h_2)\equiv
\left[\sum_{\alpha\neq\beta}^N
\frac{P_{\text{G}}S_\alpha^zS_\beta^zP_{\text{G}}}
{\vert\eta_\alpha -\eta_\beta\vert^2}\varphi_{mn}\right]\!
(z_1,\ldots, z_M;h_1,h_2)\nonumber \\
&=&\left\{\sum_{i\neq j}^{M}\frac{1}{|z_{i}-z_{j}|^{2}}+
\sum_{i=1}^{M}\frac{1}{|z_{i}-h_1|^{2}}+\sum_{i=1}^{M}
\frac{1}{|z_{i}-h_2|^{2}}+\frac{1/2}{\vert h_1 -h_2 \vert^2}\right.
\nonumber \\
& &\phantom{,,}-N\frac{N^2-1}{48}\Bigg\}\varphi_{mn}.
\end{eqnarray}

The potential term yields 
\begin{eqnarray}
& &\hspace{-10mm}\Big[H_{\text{V}}\varphi_{mn}\Big]
(z_{1},\ldots,z_{M};h_1,h_2)\equiv
\left[ \sum_{\alpha\neq\beta}^N
\frac{P_{\text{G}}
\left(-\frac{1}{4}n_\alpha n_\beta+n_\alpha-\frac{1}{2}\right)
P_{\text{G}}}{\vert \eta_\alpha - \eta_\beta \vert^2}
\varphi_{mn}\right]\!(z_{1},\ldots,z_{M};h_1,h_2)\nonumber \\
&=&\left\{-\frac{1}{2}\frac{1}{\vert h_1-h_2 \vert^2}-\frac{N^2-1}{12}+
\frac{N}{4}\frac{N^2-1}{12} \right\}\varphi_{mn}.
\label{su2twopotential}
\end{eqnarray}
 
The charge kinetic terms deserve particular care as new techniques 
are required.  
  
For the $\uparrow$-charge kinetic term, we first observe that
$[c_{\beta\uparrow}^{\phantom{\dagger}}c_{\alpha\uparrow}^{\dagger}\varphi_{mn}
] (z_1 , ... , z_{M};h_1,h_2)$ is identically zero unless one of the
arguments $z_1 , ... , z_{M}$ equals $\eta_\alpha$. We thus find
\begin{eqnarray}
& &\hspace{-10mm}\Big[H_{\text{C}}^{\uparrow}\varphi_{mn}\Big]
(z_{1},\ldots,z_{M};h_1,h_2)\equiv
\left[ \sum_{\alpha\neq\beta}^N
\frac{P_{\text{G}}c_{\beta\uparrow}^{\phantom{\dagger}}
c_{\alpha\uparrow}^{\dagger}P_{\text{G}}}{\vert \eta_\alpha - 
\eta_\beta \vert^2}\varphi_{mn}\right]\!
(z_{1},\ldots,z_{M};h_1,h_2)\nonumber \\
&=&\sum_{\alpha=h_1,h_2}\sum_{\beta \neq \alpha}^N
\frac{1}{\vert \eta_{\alpha}-\eta_\beta\vert^2}
\varphi_{mn}\nonumber \\
&=&\sum_{\beta \neq h_2}^{N}\frac{\varphi_{mn}
(z_1,\dots,z_M;h_1,\eta_{\beta})}{\vert h_2-\eta_{\beta} \vert^2}+
\sum_{\beta \neq h_1}^{N}\frac{\varphi_{mn}(z_1,\dots,z_M;\eta_{\beta},h_2)}
{\vert h_1-\eta_{\beta} \vert^2}\nonumber \\
&=& \sum_{l = 0}^{M}\sum_{\beta \neq h_2}^N
\frac{\eta_\beta^{n}(\eta_\beta - h_2)^l}
{l!\vert h_2 - \eta_\beta\vert^2}
\frac{\partial^l}{\partial \eta_{\beta}^l}
\left( \frac{ \varphi_{mn}(z_{1},\ldots,z_{M};h_1,\eta_{\beta})}
{\eta_{\beta}^{n}} \right)\rule[-20pt]{0.3pt}{36pt}_{\;\eta_{\beta}=h_2} 
\nonumber \\
& &+\sum_{l = 0}^{M}\sum_{\beta \neq h_1}^N
\frac{\eta_\beta^{m}(\eta_\beta - h_1)^l}
{l!\vert h_1 - \eta_\beta\vert^2}
\frac{\partial^l}{\partial \eta_{\beta}^l}
\left( \frac{ \varphi_{mn}(z_{1},\ldots,z_{M};\eta_{\beta},h_2)}
{\eta_{\beta}^{m}} \right)
\rule[-20pt]{0.3pt}{36pt}_{\;\eta_{\beta}=h_1}\nonumber \\
&=&\sum_{l=0}^{M}\frac{h_1^{m+l}}{l !}B_l^m \frac{\partial^l}{\partial h_1}\left(\frac{\varphi_{mn}}{h_1^m}\right)+\sum_{l=0}^{M}\frac{h_2^{n+l}}{l !}B_l^n \frac{\partial^l}{\partial h_2}\left(\frac{\varphi_{mn}}{h_2^n}\right) \nonumber \\
&=&\left\{ \left(\frac{N^2-1}{6}+\frac{m(m-N)}{2}+\frac{n(n-N)}{2}\right)
h_1^{m}h_2^n\right.\nonumber \\
& &\phantom{,}-\left(\frac{N-1}{2}-m\right)h_1^{m+1}h_2^n
\frac{\partial}{\partial h_1}-\left(\frac{N-1}{2}-n \right)h_1^m h_2^{n+1}
\frac{\partial}{\partial h_2}\nonumber \\*
& &\left.\phantom{}+\frac{1}{2} h_1^{m+2}h_2^n
\frac{\partial^{2}}{\partial h_1^{2}} +\frac{1}{2}h_1^mh_2^{n+2}
\frac{\partial^2}{\partial h_2^2}\right\} 
\frac{\varphi_{mn}(z_{1},\ldots,z_{M};h_1,h_2)}{h_1^{m}h_2^n}, \label{twochargehole}
\end{eqnarray}
where we have introduced the coefficients 
$B_l^{n}= -\sum_{\beta =1}^{N-1}\eta_{\beta}^{n+1}
(\eta_{\beta}-1)^{l-2}$, which are evaluated in Appendix~\ref{asecc}. 
\eqref{twochargehole} is valid if and only if
$0\leq n,m \leq (N+2)/2$, which finally leads to the restriction
\eqref{restriction} for the actual $\Psi_{mn}^{\text{ho}}$'s.

For the treatment of the $\downarrow$-charge kinetic term we avail
ourselves of the fact that $\varphi_{mn}$ can be equally expressed by
the up-spin or down-spin variables, as we show in
Appendix~\ref{asecw}. If we denote the down-spin coordinates by $w_i$,
we obtain
\begin{eqnarray}
& &\hspace{-10mm}\Big[H_{\text{C}}^{\downarrow}\varphi_{mn}\Big]
(z_{1},\ldots,z_{M};h_1,h_2)\equiv
\left[ \sum_{\alpha\neq\beta}^N
\frac{P_{\text{G}}c_{\beta\downarrow}^{\phantom{\dagger}}
c_{\alpha\downarrow}^{\dagger}P_{\text{G}}}{\vert \eta_\alpha - 
\eta_\beta \vert^2}
\varphi_{mn}\right]\!(z_{1},\ldots,z_{M};h_1,h_2)\nonumber \\
&=&\left\{ \left(\frac{N^2-1}{6}+\frac{m(m-N)}{2}+\frac{n(n-N)}{2}\right)
h_1^{m}h_2^n\right.\nonumber \\
& &\phantom{,}-\left(\frac{N-1}{2}-m\right)h_1^{m+1}h_2^n
\frac{\partial}{\partial h_1}-\left(\frac{N-1}{2}-n \right)h_1^m h_2^{n+1}
\frac{\partial}{\partial h_2}\nonumber \\*
& &\left.\phantom{,}+\frac{1}{2} h_1^{m+2}h_2^n
\frac{\partial^{2}}{\partial h_1^{2}} +\frac{1}{2}h_1^mh_2^{n+2}
\frac{\partial^2}{\partial h_2^2}\right\} 
\frac{\varphi_{mn}(w_{1},\ldots,w_{M};h_1,h_2)}{h_1^{m}h_2^n}. 
\end{eqnarray}
Using identities verified in Appendix~\ref{asecd} for the derivatives with
respect to the $z_i$'s and $h_{1,2}$'s, the total charge-kinetic term
becomes 
\begin{eqnarray}
& &\hspace{-10mm}\left[ \sum_{\alpha\neq\beta}^N
\frac{ H_{\text{C}}^{\downarrow}+H_{\text{C}}^{\uparrow}}{\vert \eta_\alpha - 
\eta_\beta \vert^2} 
\varphi_{mn}\right]\!(z_{1},\ldots,z_{M};h_1,h_2)\nonumber \\
&=&\left\{\left[\frac{N^2-1}{3}+m(m-N)+n(n-N)-C_2-C_1^2+m\left(C_1-
\frac{1}{2}\right)\right.\right.\nonumber \\
& &\left.\phantom{,}+n\left(C_1-\frac{1}{2}\right)-\frac{h_1+h_2}{h_1-h_2}
\left(\frac{m-n}{2}\right)\right]h_1^mh_2^n+h_1^{m+2}h_2^{n}
\frac{\partial^2}{\partial h_1^2}+h_1^mh_2^{n+2}\frac{\partial^2}
{\partial h_2^2}\nonumber \\
& &\phantom{,}+h_1^mh_2^{n+1}\frac{h_2}{(h_1-h_2)}
\frac{\partial}{\partial h_2}+h_1^{m+1}h_1^n\frac{h_1}{h_2-h_1}
\frac{\partial}{\partial h_1}+C_1h_1^{m+1}h_2^n\frac{\partial}{\partial h_1}\nonumber \\
& &\phantom{,}+C_1h_1^mh_2^{n+1}\frac{\partial}{\partial h_2}+h_1^mh_2^n
\sum_{i}^M\frac{h_2^2}{(z_i-h_2)^2}+h_1^mh_2^n\sum_{i}^M
\frac{h_1^2}{(z_i-h_1)^2}\nonumber \\
& &\left.\phantom{,}+\frac{h_1^2+h_2^2}{(h_1-h_2)^2}\right\}
\frac{\varphi_{mn}(z_{1},\ldots,z_{M};h_1,h_2)}{h_1^mh_2^n} , 
\end{eqnarray}
with the constants
$C_1=\sum_{\alpha=1}^{N-1}1/(1-\eta_{\alpha})=(N-1)/2$ and
$C_2=\sum_{\alpha=1}^{N-1}1/(1-\eta_{\alpha})^{2}=(6N-5-N^2)/12$
introduced and evaluated in~\cite{BGLtwospinon}. Summing up all terms,
we finally obtain the action of the Hamiltonian \eqref{ky2} on the
auxiliary wave functions $\varphi_{mn}$:
\begin{eqnarray}
\hspace{-8mm}H_{\text{KY}}\varphi_{mn}
&=&\frac{2\pi^2}{N^2}\left\{\frac{8-9N}{8}+m(m-N)+n(n-N)+m\frac{N-2}{2}+n
\frac{N-2}{2}\right.\nonumber \\
& &\left.-\frac{1}{2}\frac{h_1+h_2}{h_1-h_2}(m-n)+
\frac{h_1^2+h_2^2}{(h_1-h_2)^2}\right\}\varphi_{mn}. 
\label{dummysu2two}
\end{eqnarray}

With \eqref{eq:auxzerlegung} this implies
\begin{eqnarray}
\hspace{-8mm}H_{\text{KY}}\Psi_{mn}^{\text{ho}}&=&\frac{2\pi^2}{N^2}\left[\Bigg(-\frac{8+N}{8}+1+\left(m-\frac{N}{2}\right)m+
\left(n-\frac{N}{2}\right)n\Bigg)\Psi_{mn}^{\text{ho}}\right.\nonumber \\
& &+\frac{m-n}{2}(h_1-h_2)\frac{h_1+h_2}{h_1-h_2}(h_1^mh_2^n-h_1^nh_2^m)
\Psi_0\Bigg] \nonumber \\
&=&\frac{2\pi^2}{N^2}\left[-\frac{N}{8}+\left(m-\frac{N}{2}\right)m+
\left(n-\frac{N}{2}\right)n
+\frac{m-n}{2}\right]\Psi_{mn}^{\text{ho}}\nonumber\\
& &+\frac{2\pi^2}{N^2}(m-n)\sum_{l=1}^{\lfloor \frac{m-n}{2}\rfloor}
\Psi_{m-l,n+l}^{\text{ho}}, \label{twoholonenergy1}
\end{eqnarray}
where we have used $\frac{x + y}{x - y} (x^{m} y^{n} - x^{n} y^{m}) =
2 \sum_{l = 0}^{m-n} x^{m-l} y^{n+l} - (x^{m} y^{n} + x^{n} y^{m})$.
The symbol $\lfloor \; \rfloor$ denotes the floor function, \ie
$\lfloor x\rfloor$ is the largest integer $l\le x$. First, note that
the action of the Hamiltonian on the $\Psi_{mn}^{\text{ho}}$'s is
trigonal, \ie the ``scattering'' in the last line is only to lower
values of $m-n$. Second, \eqref{twoholonenergy1} shows that the states
$\Psi_{mn}^{\text{ho}}$ form a non-orthogonal set. We will now proceed
to construct an orthogonal basis of energy eigenfunctions.

\subsection{Energy eigenstates}

Due to the trigonal structure of the Hamiltonian when acting on the
$\Psi_{mn}^{\text{ho}}$'s we can derive the energy eigenstates using
the Ansatz
\begin{equation}
\ket{\Phi_{mn}^{\text{ho}}}=\sum_{l=0}^{\lfloor\frac{m-n}{2}\rfloor}a_{l}^{mn}
\ket{\Psi_{m - l \; n + l}^{\text{ho}}}, 
\label{su2twoenergystates}
\end{equation}
which yields the recursion relation
\begin{equation}
a_{l}^{mn}=-\frac{1}{2l(l-\frac{1}{2}+n-m)}\sum_{k=0}^{l-1}a_{k}^{mn}(m-n-2k),
\quad a_0^{mn}=1. \label{su2holonrecursion}
\end{equation}
This defines the two-holon energy eigenstates
(\ref{su2twoenergystates}). The energies are given by
\begin{equation}
E_{mn}^{\text{ho}}=E_0+\frac{2\pi^2}{N^2}\bigg[\left(m-\frac{N}{2}\right)m
+\left(n-\frac{N}{2}\right)n+\frac{m-n}{2}\biggr],
\label{su2twofinalenergy} 
\end{equation}
where the momentum quantum numbers satisfy 
\begin{equation}
0 \leq n \leq m \leq \frac{N}{2}, \label{drestrict}
\end{equation}
and the total momentum is given by \eqref{su2twomomentum}.

For the lowest energy state,~\eqref{su2twofinalenergy} simplifies (up
to an additive constant $\pi^2/12N$) to the ground-state energy of the
chain doped with two holes, which is a special case of the result by
Kuramoto and Yokoyama~\cite{KuramotoYokoyama} for the ground state at
general filling fraction.

\section{Fractional statistics}\label{secfra}

Fractional statistics in one-dimensional systems was originally
introduced by Haldane~\cite{Haldane91prl2} in the context of
non-trivial state counting rules.  Recently, it was realized that the
fractional statistics of spinons in the HSM also manifests itself in
specific quantization rules for the individual spinon
momenta~\cite{GreiterS05,Greiter}.  We now apply this line of analysis
to the holon excitations in the KYM.

To begin with, let us recall that the asymptotic Bethe ansatz
solution of the KYM~\cite{Haldane94,essler95prb13357} implies that the
holons are free, \ie that they interact only through their statistics,
while there is no position or momentum dependent interaction potential
between them.  This induces us to rewrite the two-holon
energy~\eqref{su2twofinalenergy} as
\begin{equation}
E_{mn}^{\text{ho}}=E_0+\epsilon^{\text{ho}}(p_m)+\epsilon^{\text{ho}}(p_n), 
\label{fractional}
\end{equation}
where we assume the one-holon dispersion \eqref{1hdisp} and introduce 
the single-holon momenta according to
\begin{equation}
p_m=-\frac{\pi}{2}+\frac{2\pi}{N}\left(m+\frac{1}{4}\right), \quad 
\quad p_n=-\frac{\pi}{2}+\frac{2\pi}{N}\left(n-\frac{1}{4} \right).
\label{pmpn}
\end{equation} 
Note that the fractional shifts of $\frac{2\pi}{N}\cdot\frac{1}{4}$ in
$p_m$ and $p_n$ occur in opposite directions.  Since $n\le m$, the
momenta are shifted away from each other, implying $p_m>p_n$.  The
shifts directly follow from \eqref{fractional}; any other assignment
of the single particle momenta would yield an additional interaction
term in the energy, corresponding to the last term in
\eqref{su2twofinalenergy} above.  For the difference in the individual
holon momenta we hence obtain
\begin{equation}
  p_m-p_n=\frac{2\pi}{N}\left(\frac{1}{2}+\text{integer}\right).
\label{eq:fracstat}
\end{equation}
We interpret this result as a direct manifestation of the half-Fermi
statistics of the holons, as the shift in the single particle momenta
can be attributed to a statistical phase acquired by the states as the
holons pass through each other~\cite{Greiter}.  Indications of the
half-Fermi statistics of the holons have previously been observed in
thermodynamical quantities~\cite{Katokura,Kurakato} of the KYM as well
as the electron addition spectrum~\cite{Arikawa1,Arikawa2}.

Let us now elaborate on the general implictions of this result.  The
wave functions we have obtained above are of course eigenstates of the
KYM only, which is as idealized as integrable and exactly soluble
models tend to be.  The quantization rules for the single particle
momenta we have obtained for this model, however, have a much broader
validity.  As mentioned above, the unique feature of the KYM is that
the holons are free in the sense that they only interact through their
fractional statistics.  The single particle momenta of the holons are
hence good quantum numbers, which assume fractionally spaced values.
For two holons, these are given by \eqref{pmpn}.  The crucial
observation in this context is that the statistics of the holons is a
quantum invariant and as such independent of the details of the
model.  This implies directly that the fractional spacings are of
universal validity as well.  If we were to supplement the model we
have studied by a potential interaction between the holons, say a
Coulomb potential, this interaction would introduce scattering matrix
elements between the exact eigenstates we obtained and labeled
according to their fractionally spaced single particle momenta.  These
momenta would hence no longer constitute good quantum numbers.  The
new eigenstates would be superpositions of states with different
single particle momenta, which individually, however, would still
possess the fractionally shifted values.  In other words, looking at
the quantization condition \eqref{eq:fracstat}, the ``1/2'' on the
left of the equation will still be a good quantum number, while the
``integer'' will turn into a ``superposition of integers'' in the
presence of an interaction between the holons.

\section{Conclusions}\label{seccon}
In this article we have studied the two-holon states of the Kuramoto--Yokoyama
model. We constructed the explicit two-holon wave functions and derived their
momenta and energies. The results display the half-Fermi statistics of the
holons, which manifests itself in a shift of $\frac{1}{2}\frac{2\pi}{N}$ in
the difference of the individual holon momenta.

\section*{ACKNOWLEDGMENTS}

RT was supported by a Ph.D.\ scholarship of the Studienstiftung des deutschen 
Volkes.  DS was supported by the German Research Foundation (DFG) through 
grant GK 284.

\appendix
\section{B-series}\label{asecc}

Evaluation of the series
\begin{equation}
B_l^{n}= -\sum_{\beta =1}^{N-1}\eta_{\beta}^{n+1}
(\eta_{\beta}-1)^{l-2} , 
\label{eq:app-twoseries}
\end{equation}
with $l$ restricted to $0 \leq l \leq M = (N-2)/2$ yields
\begin{equation}
B_0^n=\frac{N^2-1}{12}+\left(\frac{n(n-N)}{2}\right) \quad\mathrm{for}\quad 0 
\le n \le N,
\label{eq:app-twoseries1}
\end{equation}
\begin{equation}
B_1^n=\left \{ 
  \begin{array}{lll}
    \displaystyle n-\frac{N-1}{2} \quad&\mathrm{for} &0 \le n <N, \\[10pt] 
    \displaystyle -\frac{N-1}{2}  \quad&\mathrm{for}  &n=N,
  \end{array} \right.
\label{eq:app-twoseries2}
\end{equation}
\begin{equation}
B_2^n=\left \{ 
  \begin{array}{lll}
   1  \quad   &\mathrm{for} &0 \le n \le N-2,\ n=N, \\
   1-N  \quad &\mathrm{for} &n=N-1,
  \end{array} \right.
\label{eq:app-twoseries3}
\end{equation}
\begin{equation}
B_l^n=\left \{
  \begin{array}{lll}
    0 \quad                                   &\mathrm{for}&\displaystyle 
    l \ge 3, \ 0 \le n \le \frac{N+2}{2} , \\[10pt]
    \displaystyle N{l-2 \choose N-n-1}\quad  &\mathrm{for}&\displaystyle 
    l \ge 3, \ \frac{N+2}{2} < n \le N.
  \end{array} \right.
 \label{eq:app-twoseries4}
\end{equation}
{\em Proof:} 
$B_0^n$, $B_1^n$, and $B_2^n$ are found by straight forward evaluation
of the respective sums. For (\ref{eq:app-twoseries4}) consider
\begin{eqnarray*}
B_l^n&=&-\sum_{\alpha=1}^{N-1} \eta_\alpha^{n+1}\sum_{k=0}^{l-2}
{l-2 \choose k}(-1)^{l-k-2}\eta_\alpha^k\\
&=&\sum_{k=0}^{l-2}{l-2 \choose k}(-1)^{l-k-1}
\Bigl(1-\sum_{\alpha=1}^N \eta_\alpha^{k+l+1}\Bigr)\\
&=&\left  \{
 \begin{aligned}
 -\sum_{k=0}^n {n \choose k}(-1)^{n-k}=0 \quad \mathrm{for} \quad 3 \leq l, 0\leq n \leq (N+2)/2, \\
 \sum_{k=0}^{l-2} {l-2 \choose k} N \delta_{k,N-1-n}=
N{l-2 \choose N-n-1} \quad \mathrm{otherwise.}
 \end{aligned} \right.
\end{eqnarray*}
For the last steps note that the the binomial coefficients of even
and odd sites equal each other.

\section{Wave function in $\uparrow$- and $\downarrow$-spin
  coordinates}
\label{asecw}

The wave functions $\Psi_{mn}^{\text{ho}}$ can be equally expressed
either in up-spin ($z$) or down-spin coordinates ($w$):
\begin{eqnarray}
& &\hspace{-8mm}\frac{\Psi_{mn}^{\text{ho}}(z_{1},...,z_{M};h_1,h_2)}{h_1^{m}h_2^{n}
(h_1-h_2)}\nonumber \\
&=&(-1)^{\frac{1}{2}M(M+1)}\frac{\prod_{j}^M (h_1-z_{j})(h_2-z_j)z_j
\prod_{i\neq j}^{M}(z_{i}-z_{j})\prod_{l, j}^{M}(w_{l}-z_{j})}{\prod_{l,
  j}^{M}
(w_{l}-z_{j})}\nonumber\\
&=&\frac{\Psi_{mn}^{\text{ho}}(w_{1},...,w_{M};h_1,h_2)}{h_1^{m}h_2^{n}
(h_1-h_2)} . \label{varident}
\end{eqnarray}
This identity applies to the auxiliary wave functions $\varphi_{mn}$, as
the prepolynomial contains only the coordinates $h_{1,2}$.

\section{A derivative identity}\label{asecd}

The necessary relation for the $\downarrow$-charge kinetic term is
\begin{eqnarray}
& &\hspace{-8mm}\sum_{i\neq j}\frac{h_2^2}{(w_i-h_2)(w_j-h_2)}\nonumber\\
&=&-C_2+C_1^2+2\sum_{i}\frac{h_2^2}{(z_i-h_2)^2}+\sum_{i\neq j}
\frac{h_2}{z_i-h_2}\frac{h_2}{z_j-h_2}+2\frac{h_2^2}{(h_1-h_2)^2}\nonumber\\
& &+2C_1\sum_{i}\frac{h_2}{z_i-h_2}+2C_1\frac{h_2}{h_1-h_2}+2\frac{h_2}{h_1-h_2}
\sum_{i}\frac{h_2}{z_i-h_2},\label{derident1}
\end{eqnarray}
with $C_1$ and $C_2$ defined as above.~\eqref{derident1} is also valid for $h_1 \leftrightarrow h_2$.

\end{document}